\documentclass[conference]{IEEEtran}
\IEEEoverridecommandlockouts
\usepackage{cite}
\usepackage{amsmath,amssymb,amsfonts}
\usepackage{algorithmic}
\usepackage[pdftex]{graphicx}
\usepackage{textcomp}
\usepackage{url}

\def\BibTeX{{\rm B\kern-.05em{\sc i\kern-.025em b}\kern-.08em
    T\kern-.1667em\lower.7ex\hbox{E}\kern-.125emX}}
\begin{document}

\newcommand{\Commit}{{\sf Commit}}
\newcommand{\ComOpen}{{\sf ComOpen}}
\newcommand{\com}{{\sf com}}
\newcommand{\dec}{{\sf dec}}
\newcommand{\Hash}{{\sf Hash}}
\newcommand{\C}{{\mathcal C}}

\newcommand{\Source}{{\mathcal S}}
\newcommand{\Prover}{{\mathcal P}}
\newcommand{\Verifier}{{\mathcal V}}

\newcommand{\SC}{{\mathcal SC}}

\newcommand{\vko}{{\sf vk}_{\mathcal{O}}}
\newcommand{\sigko}{{\sf sigk}_{\mathcal{O}}}

\title{A Sealed-bid Auction with Fund Binding: Preventing Maximum Bidding Price Leakage
\thanks{An extended abstract appeared at Computer Security Symposium 2021~\cite{ChinEOS21} and IEEE Blockchain 2022~\cite{ChinEOS22}.}
}


\author{
\IEEEauthorblockN{Kota Chin}
\IEEEauthorblockA{\textit{University of Tsukuba}\\\textit{National Institute of}\\\textit{Information and}\\\textit{Communications Technology}\\\textit{Japan}}\\
\and
\IEEEauthorblockN{Keita Emura}
\IEEEauthorblockA{\textit{Kanazawa University}\\\textit{National Institute of}\\\textit{Information and}\\\textit{Communications Technology}\\\textit{Japan}}\\
\and 
\IEEEauthorblockN{Kazumasa Omote}
\IEEEauthorblockA{\textit{University of Tsukuba}\\\textit{National Institute of}\\\textit{Information and}\\\textit{Communications Technology}\\\textit{Japan}}\\
\and 
\IEEEauthorblockN{Shingo Sato}
\IEEEauthorblockA{\textit{Yokohama National University}\\\textit{Japan}}\\
}

\maketitle

\begin{abstract}
In an open-bid auction, a bidder can know the budgets of other bidders. Thus, a sealed-bid auction that hides bidding prices is desirable. However, in previous sealed-bid auction protocols, it has been difficult to provide a ``fund binding'' property, which would guarantee that a bidder has funds more than or equal to the bidding price and that the funds are forcibly withdrawn when the bidder wins. Thus, such protocols are vulnerable to a false bidding. 
As a solution, many protocols employ a simple deposit method in which each bidder sends a deposit to a smart contract, which is greater than or equal to the bidding price, before the bidding phase. However, this deposit reveals the maximum bidding price, and it is preferable to hide this information. 
In this paper, we propose a sealed-bid auction protocol that provides a fund binding property.
Our protocol not only hides the bidding price and a maximum bidding price, but also provides a fund binding property, simultaneously. For hiding the maximum bidding price, we pay attention to the fact that usual Ethereum transactions and transactions for sending funds to a one-time address have the same transaction structure, and it seems that they are indistinguishable. We discuss how much bidding transactions are hidden. 
We also employ DECO (Zhang et al., CCS 2020) that proves the validity of the data to a verifier in which the data are taken from a source without showing the data itself. 
Finally, we give our implementation which shows transaction fees required and compare it to a sealed-bid auction protocol employing the simple deposit method. 
\end{abstract}


\section{Introduction}
\subsection{Background}
\label{background}
Non-Fungible Tokens (NFTs) have received much attention recently, as digital assets such as artworks have been traded on blockchains, and open-bid auctions based on smart contracts have been frequently held. For example, the NFT of Kabosu (the name of the female Shiba Inu dog that became the Internet meme \lq\lq Doge") was one of the most expensive NFT of all, with a bidding price of 1696.9 ETH (approximately 400 million USD using the Ether price on June 11, 2021).%
\footnote{Transaction detail is available at \url{https://etherscan.io/tx/0x8668bb338f7cf9896db75c00e8bef18cc549d04b2dcaf1cee01dc0e1522e7e87}}
Thus, open-bid auctions based on smart contracts have been used for transactions of relatively expensive assets and have acquired a certain degree of reliability. 
They also treated relatively cheap assets; for example, the median price of NFTs traded during May 2023 was approximately 300 USD.%
\footnote{Calculation detail is available at (\url{https://dune.com/queries/351118})}
As another merit of employing smart contracts, it has a high affinity with NFTs (since they are also based on a blockchain-related technology) and it is easy to directly provide the NFT to the winner of the auction. 
%
We remark that any data preserved on the blockchain are open. Thus we need to carefully consider security and privacy when we design a protocol employing a smart contract because anyone obtains data used by the smart contract.

In an open-bid auction, there are two types, English auction and Dutch auction. In an English auction, bidders compete to raise the bidding price until  there is only one bidder left.  In a Dutch auction, the price is decreased from the initial amount until the first bidder appears.  In a sealed-bid auction, the second-price sealed-bid is called the Vickrey auction where the winner pays the second highest bid. Krishna~\cite{Krishna} showed that, based on game theory, the English auction is equivalent to the Vickrey auction when bidders evaluate the value of the item in private. Moreover, the Dutch auction is strategically equivalent to the first-price sealed-bid auction (see a nice summarization given by Bag et al.~\cite{BagHSR20}). 
Since smart contracts are recently employed, to the best of our knowledge, no game theoretic estimation against an auction based on a smart contract has been shown. However, as a fact, anyone can view the balance of addresses used during the bidding process if smart contracts are employed, and can guess the budgets of other bidders. Thus, the final selling price may be lower because bidders do not bid a price higher than the budgets of other bidders. This could be a detriment of the seller and auctioneer. Therefore, it seems desirable to guarantee that no bidder can know the balance of addresses of other bidders in advance when a smart contract is employed.

\subsection{Previous Work}

Many sealed-bid auction protocols employing smart contracts have been proposed. However, this type of protocol necessitates ensuring a bidder has funds more than or equal to the bidding price (called \lq\lq fund binding"), but because the bidding prices are hidden, this is difficult to do. Such protocols are vulnerable to a false bidding, meaning bidders indicate bidding prices, but do not have the money to be paid. 
As a countermeasure, Galal and Youssef~\cite{Galal2019TrusteeFP} and Li et al.~\cite{Li2021ABS} considered using a small deposit, wherein if bidding prices are false, then the deposit is automatically confiscated. 
However, if some bidders do not care about such a penalty because of a small deposit, then their proposals do not prevent bidders from submitting a false bidding. 
Another countermeasure against submitting a false bidding is called \lq\lq simple deposit method'' where each bidder is required to send a deposit more than or equal to the bidding price. The simple deposit method has been widely employed, e.g.,~\cite{8543821,CLXW22,KosbaMSWP16,KrolSTPR20,KadadhaMSOO20,0006VSDM21,VakiliniaBS18,abs-1901-07824,Al-SadaLA21,SharmaVSDM21,HsuM21,ConstantinidesC21}. Although each bidder can hide their actual bidding price by submitting a deposit more than the actual bidding price, the maximum bidding price is leaked because deposit information is publicly available on smart contracts. Thus, the final selling price may be lower because bidders do not bid a price higher than the highest deposit. Thus, it is preferable to hide the maximum bidding price, so items being auctioned sell closer to their actual value. 
In the Ma et al.'s sealed-bid auction protocol~\cite{MaQL19}, a commitment on a deposit is computed before the bidding phase. Thus, at first sight, their protocol provides fund binding without revealing the bidding price. However, it is not guaranteed that a bidder who computes a commitment on a bidding price has funds more than or equal to the bidding price. Thus, it is difficult to provide a fund binding property in sealed-bid auction protocols. 

\subsection{Our Contribution}

In this paper, we propose a sealed-bid auction protocol providing the fund binding property. Our protocol simultaneously provides the following: 

\begin{itemize}
\item Price Hiding: It guarantees that nobody can know bidding prices of other bidders until the revealing phase. 
 Especially, the maximum bidding price is also hidden.
\smallskip
\item Fund Binding: It guarantees that a bidder has funds more than or equal to the bidding price, and the funds are forcibly withdrawn when the bidder wins. 
\end{itemize}

\noindent 
We emphasize that sealed-bid auction protocols with the simple deposit method reveal the maximum bidding price although they provide fund binding. 
Our goal in this work is to provide price hiding and fund binding simultaneously. 

\medskip
\noindent\textbf{Our Technique.} 
To achieve our goal, we pay attention to the fact that an address of the smart contract can be calculated before deploying the contract, and bidders issue a one-time address by themselves. Then, each bidder sends a transaction to own one-time address. Here, we assume that a usual Ethereum transaction and a bidding transaction to a one-time address are indistinguishable, which is a reasonable assumption and allows us to hide the maximum bidding price. For example, let the bidding phase be from May 25 to 31, 2023 (one week). Then there are 109,516 possible one-time addresses. We discuss how much bidding transactions are hidden in Section~\ref{Tran_IND}.
We remark that, by observing transactions during the bidding phase, the maximum bidding price could be guessed because it is less than or equal to the trading price during this phase. Therefore, we also assume that the actual maximum bidding price is not revealed from transactions during the bidding phase. We discuss whether this assumption is reasonable by observing actual transactions in Section~\ref{MAX_HIDE}. 

In addition to employing one-time addresses, each bidder is required to prove that enough balance is preserved on the one-time address without revealing the balance and the address. We employ DECO (DECentralized Oracle)~\cite{ZhangMMGJ20} that allows a bidder to prove the statement above in a zero-knowledge manner. In particular, it allows bidders to prove that the balance is preserved on a one-time address. 

We implement our protocol and compare it to the simple deposit method. We show that the two protocols require almost similar fee to run the protocol (the additional fee was 5.28 USD calculated by the price on June 1, 2023). 

\medskip
\noindent
\textbf{Out of Scope for Our Work.} Our protocol hides bidding prices during the bidding phase. Each bidder sends a trapdoor (a decommitment of the underlying commitment scheme), and then the auction smart contract reveals all bidding prices and decides the winner. Although some sealed-bid auction protocols consider how to decide the winner without revealing other bidding prices, we do not consider to hide it after revealing since our goal in this work is to provide price hiding and fund binding simultaneously during the bidding phase. However, from the viewpoint of privacy, one merit of such a sealed-bid auction is that losing bids are kept secret when the auctioneer declares the winner and the winning bid only. This privacy-preserving manner is effective to protect revealing unnecessary information to run an auction. We expect that our protocol can be adopted/combined with other sealed-bid auction protocol, and price hiding during the revealing phase is left as a future work. 

In our protocol, bidders open their bid commitments in an undefined order. So, one may think that bidders do not want to open the commitments if another bidder has already opened a higher bid. Actually, the information about how much they bid in this auction might be useful to other bidders in subsequent related auctions. For example, Naor et al.~\cite{NaorPS99} hide all unsuccessful bids for this reason. In our protocol, even such bidders have an incentive to send decommitments since they have no way to withdraw the funds of the one-time addresses. Thus, we assume that all bidders send their decommitments honestly, and do not consider a case when bidders do not send their decommitments. 

\subsection{Concurrent and Independent Work}

\subsubsection{FAST} 
In addition to sealed-bid auction protocols providing small or simple deposit methods~\cite{Galal2019TrusteeFP,Li2021ABS,8543821,CLXW22,KosbaMSWP16,KrolSTPR20,KadadhaMSOO20,0006VSDM21,VakiliniaBS18,abs-1901-07824,Al-SadaLA21,SharmaVSDM21,HsuM21,ConstantinidesC21}, recently, David et al.~\cite{David2021FASTFA} proposed FAST (Fair Auctions via Secret Transactions) with the same motivation as ours. That is, the deposit reveals information about the bid, and thus it should be hidden. They introduced secret deposits based on confidential transaction~\cite{Maxwell2016}. They defined deposit committee members (which are different from bidders) and a trapdoor, which can be used to reveal the value of deposits, that is distributed among $m$ deposit committee members using a proactive secret sharing scheme~\cite{David2020ALBATROSS}. There are $\ell$ rounds in total where $\ell$ is the bit length of the bids. For each round, the anonymous veto protocol~\cite{Hao2006} is run, and if a party is found to be cheating, a smart contract automatically redistributes cheaters' deposits among the honest parties, which creates the incentive for parties to behave honestly. The main difference from our protocol is complexity because FAST requires each bidder to communicate with other bidders. That is O($\ell n$) computations are required where $n$ is the number of bidders. Conversely, our protocol does not require any interaction between bidders (a bidder is required to communicate with the oracle for running DECO, and later the bidder communicates with the auction smart contract only). Moreover, FAST needs to assume that the number of corrupted deposit committee members is less than $m/2-2$, whereas we do not require such a limitation. From our perspective, FAST employs a cryptographic approach to hide and bind the deposit whereas we employ another approach where a transaction of the usual transfer of Ethereum and a transaction for sending a bidding price to a one-time address are indistinguishable. This drastically reduces the complexity of the protocol for hiding and binding deposits. 

\subsubsection{Anonymous Vickrey Auctions On Chain} 

The anonymous vickrey auction called vickrey.xyz has been launched~\cite{AnonymousVickreyAuctionsOnChain,AnonymousVickreyAuctionsOnChain_blog}. 
As in our system, the auction leverages uninitialized CREATE2 addresses. Their main motivation besides sealed-bid is to provide anonymity, or private participation because merely knowing who is participating allows you to collude off-chain to lower the final price.  Bidders send money to an uninitialized CREATE2 address instead of sending shielded money to a contract. 
Both our system and vickrey.xyz assume that other users' usual transactions and transactions to the one-time address are indistinguishable, i.e., assuming that they just look like EOA (account) transfers (See Section~\ref{Tran_IND}). 
The main difference is that our system employs DECO to prove that enough balance is preserved on the one-time address without revealing the balance and the address. This prevents bidders to change their bidding amount after seeing other bidding amount in the revealing phase. On the other hand, the vickrey.xyz auction employs the snapshotted blockhash which encodes the root of the Ethereum storage trie at that point in time. Each bidder sends a Merkle Patricia tree proof in Ethereum of the balance of their CREATE2 address during the snapshotted blockhash. 

\section{Preliminaries}

\subsection{Ethereum and Smart Contracts}
\label{Ethereum}
Ethereum~\cite{Buterin2015ANS} is a platform for decentralized applications based on blockchain technology. It allows users to create applications that run on the Ethereum Virtual Machine (EVM), called smart contracts. The results of smart contract execution are shared and agreed upon among all Ethereum nodes, making it difficult to tamper.

ETH is a native token on Ethereum and is used to pay transaction fees. The transaction fee is calculated by multiplying the amount of used gas by gas price. The amount of gas used by the transaction varies depending on the operations. Gas price also fluctuates according to the demand of the Ethereum network.
A web service called Etherscan\footnote{\url{https://etherscan.io/}} allows users to check the balance of their addresses and transaction information.

\subsection{Oracles}
\label{oracle}
Smart contracts cannot directly retrieve data outside the blockchain, such as ETH price, stock prices, weather, or election results. Therefore, if a smart contract requires the use of these data, then someone needs to input them into the smart contract from outside. An entity that inputs data from outside the blockchain to a smart contract is called an oracle.

Currently, Chainlink~\cite{Chainlink2021} is the protocol with the largest share among the oracles. Chainlink is a decentralized oracle network that aggregates data obtained by multiple oracle nodes to prevent oracle fraud. In addition, because Chainlink nodes are financially penalized if they deviate from the Chainlink protocol, they are trustworthy in terms of their behavior according to the protocol. That is, in a cryptographic manner, oracles are modeled as honest-but curious entities where they follow the protocol procedure but they may try to extract information through the execution of the protocol. 

\subsection{DECO}
\label{DECO}

In this section, we introduce DECO~\cite{ZhangMMGJ20} which allows a user to prove the possession of data in a zero-knowledge manner wherein the data come from a particular website via TLS (Transport Layer Security). 
DECO contains three entities: a data source (e.g., a particular website) $\Source$, a prover $\Prover$, and a verifier $\Verifier$. To access $\Source$, some secret value is required such as salt. Then, $\Prover$ proves that the data come from $\Source$ without showing the data, and $\Verifier$ checks the validity of the proof. The flow of DECO is explained as follows. 
$\Source$, $\Prover$, and $\Verifier$ run a three-party handshake protocol. Then, $\Source$ obtains an encryption key and a key for MAC (Message Authentication Code), denoted by $k_{\sf MAC}$, as in the TLS protocol. That is,  DECO does not require any server-side modifications or cooperation, and $\Source$ follows the unmodified TLS protocol. 
Via the three-party handshake protocol, $\Source$ and $\Prover$ share the encryption key, and $\Prover$ and $\Verifier$ share $k_{\sf MAC}$ in a secret sharing manner ($\Prover$ obtains $k_P$ and $\Verifier$ obtains $k_V$ where $ k_{\sf MAC}=k_P+k_V$). 
In this setting, with the help of $\Verifier$, $\Prover$ can make a query sent to $\Source$ via TLS. More concretely, $\Prover$ and $\Verifier$ run a MPC (Multi-Party Computation) that takes $k_P$ and $k_V$ as input, respectively. We remark that $\Verifier$ cannot observe the content of the encrypted query because $\Verifier$ does not have the encryption key. $\Prover$ generates a commitment on the query and the response from $\Source$, and sends the commitment to $\Verifier$. Then, $\Verifier$ sends $k_V$ to $\Prover$, and $\Prover$ recovers $k_{\sf MAC}$, and checks the MAC on the response from $\Source$. Finally, $\Prover$ generates a zero-knowledge proof for the query and sends it to $\Verifier$. 

The DECO website\footnote{\url{https://www.deco.works/}} states that \emph{Chainlink~\cite{Chainlink2021} plans to perform an initial PoC of DECO, with a focus on decentralized finance applications such as Mixicles}. Then, a user would be able to prove the validity of the data containing their own personal information to a Chainlink node. In the proposed scheme, we employ DECO to prove that a bidder has funds that are equal to the bidding price to an oracle. Then the bidder is regarded as $\Prover$, the oracle is regarded as $\Verifier$, and Etherscan is regarded as $\Source$ in our protocol. 

\subsection{Hash-based Commitment Scheme}
\label{hash}

In this section we introduce a hash-based commitment scheme. A commitment scheme $(\Commit,\ComOpen)$ is defined as follows. The $\Commit$ algorithm takes as input a message $M$, and outputs a commitment $\com$ and a decommitment $\dec$. The $\ComOpen$ algorithm takes as input $\com$, $\dec$, and $M$, and outputs 0 (reject) or 1 (accept). A commitment scheme is required to provide hiding where no adversary can obtain information of $M$ from $\com$, and is required to provide binding where no adversary can produce $\dec$ and $\dec^\prime$ where $\ComOpen(\com,\dec,M)=1$, $\ComOpen(\com,\dec^\prime,M^\prime)=1$, and $M\neq M^\prime$ holds. 

Next, we select the commitment scheme to be employed in our implementation. Because the smart contract runs the $\ComOpen$ algorithm, it is desirable to select an efficient scheme to reduce the gas price. Therefore, we select a hash-based commitment scheme because it is more efficient than a scheme based on an algebraic structure such as the Pedersen commitment scheme~\cite{Pedersen91}. Let $\Hash:\{0,1\}^\ast\rightarrow \{0,1\}^k$ where $k$ is a security parameter. In the $\Commit$ algorithm, choose a random $R\xleftarrow{\$}\{0,1\}^k$, compute $\com=\Hash(M||R)$, and output $\com$ and $\dec=R$. The $\ComOpen$ algorithm outputs 1 if $\com=\Hash(M||\dec)$ holds, and 0, otherwise. This hash-based scheme provides both hiding and binding if $\Hash$ is modeled as a random oracle~\cite{Pass04alternativevariants,Unruh16}. 

The commitment scheme provides either statistical hiding or statistical binding according to the range size~\cite{Pass04alternativevariants,Unruh16}. Specifically, for $\Hash:\{0,1\}^{2n}\rightarrow \{0,1\}^{\ell(n)}$, the scheme provides statistical hiding if $\ell(n)=n/8$ (Lemma 9 in~\cite{Pass04alternativevariants} and Lemma 17 in~\cite{Unruh16}), and statistical binding if $\ell(n)=4n$ (Lemma 8 in~\cite{Pass04alternativevariants}), respectively. In our implementation, we employ SHA256 as the underlying hash function because SHA256 has been prepared in a smart contract environment as a pre-compiled function, and can be efficiently run with a low gas usage. 
For our usage, the input length is 768 bits (one-time address [256 bits]+a bidding price [256 bits]+a random $R$ [256 bits]). Thus, the hash function is defined as $\Hash:\{0,1\}^{768} \rightarrow \{0,1\}^{256}$. Because our usage does not match the above cases, the commitment scheme provides computational hiding and computational binding. 

\begin{figure*}[tb]
\centering
\includegraphics[width=0.82\linewidth]{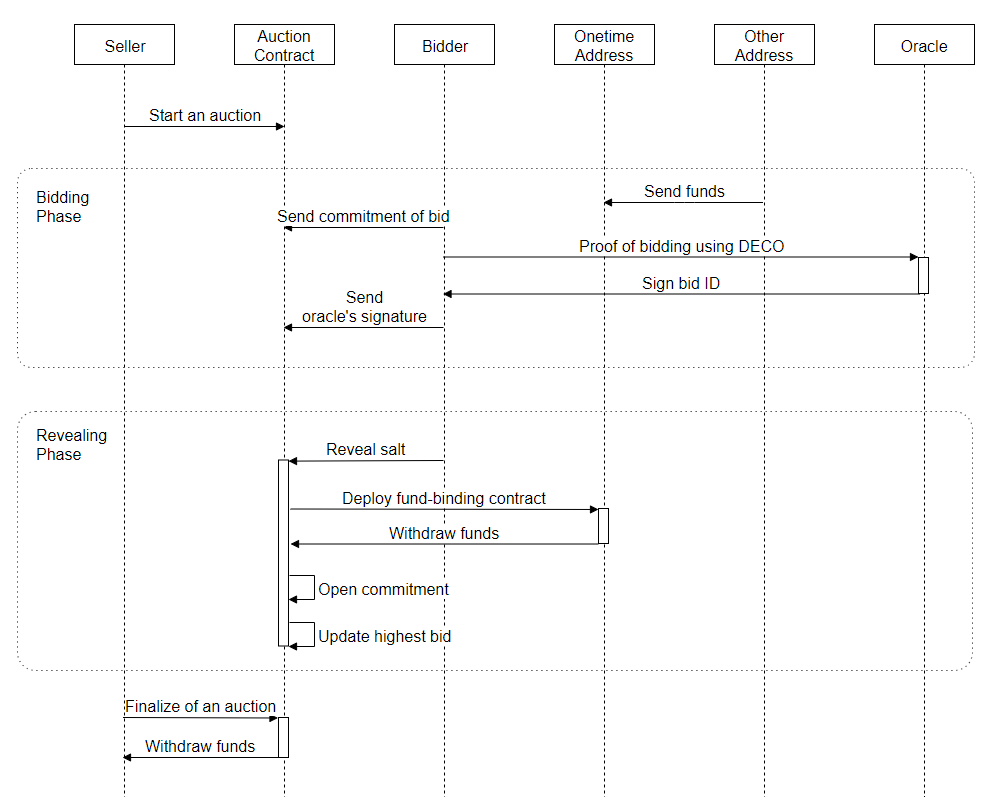}
\caption{The Sequence Diagram of the Proposed Protocol}
\label{fig:flow}
\end{figure*}

\subsection{Zero-Knowledge Proofs}
\label{ZK}

Zero-knowledge proofs/arguments for a committed value are required in DECO. In this section, we show the feasibility of the zero-knowledge argument for the hash-based commitment scheme, that is, a proof of $M$ and $R$ satisfying $\com=\Hash(M||R)$, especially when SHA256 is chosen as the underlying hash function. First, we generate an arithmetic circuit $C$ to compute SHA256 that can be obtained by the circuit generator~\cite{ParnoHG013} with a C program to compute SHA256. Next, we can produce a desirable argument system by employing zk-SNARK (zero-knowledge Succinct Non-interactive ARgument of Knowledge) for arithmetic circuits~\cite{Ben-SassonCTV14}. Ben-Sasson et al.~\cite{Ben-SassonCG0MTV14} provided a hand-optimized arithmetic circuit for the SHA256 compression function and then reduced the number of gates compared to that generated by the circuit generator~\cite{ParnoHG013}. Although we may also be able to reduce the number of gates by providing a hand-optimized arithmetic circuit for SHA256, here we only show the feasibility because zero-knowledge proofs/arguments are run in an off-chain and we focus on how to reduce the gas price for running smart contracts. 

\section{Proposed Sealed-bid Auction with Fund Binding}

In this section, we introduce the proposed protocol. 
Figure \ref{fig:flow} shows a sequence diagram of the proposed protocol.
We employ a smart contract as an auctioneer that collects bid amounts in a hidden manner and decides the winner of the auction. We refer to the smart contract as the auction contract.
The proposed protocol does not require a deposit because it reveals the maximum bidding price; thus, bidders do not directly send their own bidding price to the auction contract. 
The bidder sends funds to a fund-binding contract deployed from the auction contract.
Here, Ethereum offers CREATE2 opcode for deploying smart contracts. With the CREATE2 opcode, the bidder can compute the address of a fund-binding contract before its deployment, based on some information (See Section~\ref{IssuingOTA}), though usually the address of a smart contract is generated when the contract is deployed. We call the address of a fund-binding contract a one-time address.
In the proposed protocol, one-time addresses are used before deploying the fund-binding contract.
We assume that other users' usual transactions and transactions to the one-time address are indistinguishable. 
Then, a one-time address and its balance will be kept secret.
Next, the bidder enters a commitment of the bidding price into the auction contract, and uses DECO~\cite{ZhangMMGJ20} to prove that the balance of the one-time address is the same as the bidding price. 
We emphasize that DECO is run in an off-chain among Etherscan, a bidder, and an oracle node, and the auction contract just checks the validity of signatures sent from the oracle. 
Finally, the auction contract deploys fund-binding contracts whose addresses are bidder's one-time addresses.
Then, the auction contract withdraws the funds from one-time addresses, and determines the winning bidder by the opening result of commitments. As mentioned before, we assume that all bidders send their decommitments honestly because they have no way to withdraw the funds of the one-time addresses. 


\subsection{Starting an Auction}
A seller who wants to start an auction sends auction information such as the duration of the bidding phase and the revealing phase to the smart contract. 
In this paper, we refer to this smart contract as an auction contract. 
At this time, a unique number is assigned to each auction managed by the auction contract as an \lq\lq auction ID''. 
In our implementation, the auction smart contract incremented an \lq\lq auction ID" by 1 sequentially as each auction started. 
Since the auction contract starts each auction serially, we can assume that the ID is unique. 

\subsection{Bidding Phase}
A bidder, who wants to bid on the auction, performs the following steps during the bidding phase.

\medskip
\subsubsection{Issuing One-time Address}\label{IssuingOTA}
A one-time address is a disposable address that is issued for a certain purpose. For example, in our proposed protocol, a one-time address is used to bind the funds. If there is a trusted third party, the third party can issue a one-time address. However, assuming a trusted party is a much stronger assumption, and it would be better to avoid introducing it in cryptographic protocols as much as possible. To issue a one-time address without any trusted third party, we observe that an address of a smart contract can be calculated in advance, that is, before the smart contract is deployed.%
\footnote{This fact has been utilized in protocols such as Argent~\cite{Argent} on Ethereum.}
Specifically, we can precompute the address from the following arguments using CREATE2 opcode before deploying fund-binding contacts. 

\begin{itemize}
\item A salt
\item A bytecode of the fund-binding contact
\item An address of the auction contract
\end{itemize}

\noindent
We remark that, in our system, first the auction contract is deployed and the source code is published on platforms like Etherscan for providing transparency. The auction contract will deploy the fund-binding contract to withdraw the bidding prices and thus the auction contract incorporates the source code of the fund-binding contract. Thus, when one compiles the source code, then one obtains the bytecode of the fund-binding contract (in addition to the bytecode of the auction contract). 
In this paper, the salt is defined as a concatenation of the auction ID, the bidder's address, which is used to deposit to the one-time address, and a random value chosen by the bidder. 
Thus, the bidder is the only one who can know the one-time address until the salt is revealed.
Note that only an auction contract can deploy a fund-binding contract to a one-time address and can withdraw funds from a one-time address. 

\medskip
\subsubsection{Bidding}
A bidder sends funds to the one-time address issued.
A bidder is required to send the bidding amount to a one-time address from an address that the bidder does not use to send bidding information to the auction contract. 
If the bidder uses the same address, it helps to distinguish whether an address is related to the auction, and our assumption does not hold.

\medskip
\subsubsection{Proving of Bidding using DECO}
\label{ProvingOfBidding}
Each bidder is required to prove that the bidding price is preserved at a one-time address to provide the fund-binding property. 
In the proposed protocol, the bidder proves to the oracle node that the balance of the one-time address displayed by Etherscan is equal to the bidding price using DECO.%
\footnote{We modified an application of binary options using DECO~\cite{ZhangMMGJ20}.}
Let Etherscan be $\Source$, the bidder be $\Prover$, the oracle be $\Verifier$, and $(\vko,\sigko)$ be the oracle's verification key and signing key.
The secret information for the bidder's access to Etherscan is the one-time address $\theta_p$. 
First, the bidder sets $P$ as the bidding price, calculates $(\com_P,\dec_P)=\Commit(P)$, and sends $\com_P$ to the auction contract.
The auction contract issues a unique number as the bid ID that corresponds to the commitment. 
Next, the Etherscan, a bidder, and the oracle execute DECO. 
The bidder accesses the page of Etherscan that describes the balance value of the one-time address,  and obtains the balance value $P^\ast$. 
Note that the oracle does not know which page in the Etherscan corresponds to the bidder because the query is hidden.
This means that the oracle cannot learn the one-time address of the bidder.
The bidder then provides the oracle with a zero-knowledge proof of knowledge that $P^*$ was obtained from $\theta_p$, plus
${\sf ZK}\text{-}{\sf PoK}\{P^*: \com_P=\Commit(P^*)\}$. 
This shows that the committed value P stored in the auction contract before the execution of DECO is $P^*$.
If the proof is valid, the oracle generates a signature on the bid ID using $\sigko$ as a credential.
Finally, the bidder sends the oracle's signature to the auction contract, and if the signature is valid, the auction contract accepts the bidding of $\com_P$ specified by the bid ID.

\medskip
\subsection{Revealing Phase}
In the revealing phase, the bidder reveals the bidding price according to the following procedure.

\medskip
\subsubsection{Revealing Salt and Decommitment}
The bidder sends the auction ID, the bidding price, the salt, and the decommitment to the auction contract. 
Based on the salt, the auction contract deploys a fund-binding contract whose address has been generated by the bidder using CREATE2 opcode. 
We remark that, a fund-binding contract is allowed to transfer funds to the auction contract in our implementation, and thus each bidder cannot withdraw the balance without the auction contract. 
After the funds is transferred from the fund-binding contract to the auction contract, the auction contract runs the $\ComOpen$ algorithm (described in Section \ref{hash}) against commitments whose bid ID have been verified. 
The auction contract obtains a bidding price, and if it is greater than the current highest bid, then the auction contract updates the current highest bid. 
Otherwise, if it is less than the current highest bid, then the auction contact refunds the funds. We assume that there is one highest bid and do not consider the case that there are many highest bids because it depends on a rule of the auction. 
We remark that the auction contract only holds the funds of the highest bidder in the first-price auction. However, our system can handle the second-price auction easily where the auction contract preserves the second-highest bid when it updates the current highest bid. 

Since funds of the one-time contract have been verified using DECO at the bidding phase, funds will never be less than the committed amount. 
One may wonder the freshness of the data source. Actually, a bidder can send ETH to the one-time address even after proving of the bidding price using DECO. 
Even then, we assume that the auction contract selects the committed bidding price. On the other hand, if the auction contract employs the funds withdrawn from the one-time contract, then this rule gives the bidder a room for increasing the funds after the bidding phase. 

We remark that a bidder may bid at multiple prices and open only the lowest bid that wins the auction during the revealing phase. Since deploying the fund-binding contract is the only way to withdraw funds from a one-time address, only the auction contract can withdraw the funds. This means that any funds linked to unrevealed bids stay locked, and the bidder needs to open all bids to withdraw their own funds. Thus the highest bid of the multiple bids is selected as the bidder's bid. Thus, there is no benefit that bidders send multiple bids in our system. 

\medskip
\subsection{Finalization of an Auction}
The seller finalizes the auction by using the auction contract's finalize function. 
The auction contract determines the winning bidder on the highest bids in the revealing phase, and the seller receives the funds.
As a special case, if the auctioned item is an NFT, the auction contract directly transfers the NFT to the winning bidder right after the auction contract withdraws the bidding price.\footnote{Unlike to NFT, if a smart contact does not treat an auctioned item, then there is room for argument on whether exchange between the funds and the auctioned item is honestly run. Thus, it seems reasonable to employ fair exchange protocols, especially smart contract-based protocols~\cite{AvizhehHS22,EckeyFS20,DziembowskiEF18,LinHHC21} in addition to our protocol.}

\subsection{Security Discussion}

We assume that all parties follow the protocol description (i.e., semi-honest parties), and we do not introduce any trusted third party in our protocol. Then, our protocol provides price hiding if the underlying commitment scheme provides hiding, the DECO protocol provides zero-knowledge for hiding balances, and our two assumptions hold: (1) bidding transactions to one-time addresses and usual transactions are indistinguishable, and (2) the maximum bidding price is not revealed for transactions during the bidding phase.
We discuss how much these assumptions are reasonable in Sections~\ref{Tran_IND} and \ref{MAX_HIDE}. 

Moreover, our protocol provides fund binding if the underlying commitment scheme provides binding, the underlying signature scheme (run by the oracle) is existentially unforgeable under chosen-message attack (EUF-CMA), and the DECO protocol provides soundness for proving balances. Here, soundness means that no proof for false statements (e.g., a balance is less than and not equal to the bidding price) is accepted. 

\section{Implementation}
\subsection{Gas Prices for Running the Proposed Protocol}
\label{estimate_gas}
In this section, we discuss the fees of operating auctions.
As mentioned in Section \ref{Ethereum}, we need to pay a fee called ``gas'' to operate a smart contract. 
The gas and USD-denominated fees for bidders' processing in the auction of the proposed protocol are shown in Table \ref{gascost_bidder}. This paper assumes that the gas price is 39 gwei (median gas price on June 1, 2023\footnote{\url{https://dune.xyz/queries/4294/11099}}) and 1 ETH is 1900 USD (Approximate price on June 1, 2023\footnote{\url{https://www.coingecko.com/en/coins/ethereum}}). 
Normally, a bidder will perform each operation shown in Table \ref{gascost_bidder} one by one at a time, so the bidder will consume about 19.65 USD worth of ETH in a series of operations.
The amount of gas consumed varies depending on the length of the salt. However it is considered within the margin of error due to the high volatility of gas price and ETH prices.
In addition, the fee for the series of operations is independent of the number of bidders.

\begin{table}[h]
\centering
\caption{Cost of Gas for Bidder's Operation (Proposed Protocol)}
\label{gascost_bidder}
\begin{tabular}{lll}
\hline
Operation & Used Gas & Fee \\ \hline
Sending funds     & 21,000   & 1.56 USD                         \\
Committing bid    & 68,903   & 5.10 USD                         \\
Proving bid  & 52,755   & 3.90 USD                         \\
Revealing bid       & 122,546  & 9.08 USD                        \\ \hline
\end{tabular}
\end{table}

Next, the gas and fees for the seller's operation are shown in Table \ref{gascost_seller}.
The gas is approximately 15.32 USD, although it varies depending on the length of the information entered into the smart contract at the start of the auction.
For example, assuming an NFT auction, in addition to the gas shown in Table \ref{gascost_seller}, additional gas required to send the NFT is also necessary.

\begin{table}[h]
\centering
\caption{Cost of Gas for Seller's Operation (Proposed Protocol)}
\label{gascost_seller}
\begin{tabular}{lll}
\hline
Operation           & Used Gas & Fee  \\ \hline
Starting auction & 166,510  & 12.33 USD                         \\
Finalization auction  & 40,312   & 2.99 USD                         \\ \hline
\end{tabular}
\end{table}

\subsection{Comparison}
First, we consider the simple deposit method.
As described in Section \ref{background}, this method is problematic in that the maximum bidding price is exposed.
Therefore, we can examine the fee for hiding the maximum bidding price by comparing the proposed protocol with the simple deposit method.
Because the fee depends on the amount of data stored in the smart contract and the amount of computation, we implemented the simple deposit method on a smart contract implemented in the proposed protocol without changing the structure of the smart contract as much as possible. 
Table \ref{gascost_bidder_naive} shows the amount of used gas and fee for operations performed by the bidder. 

\begin{table}[h]
\centering
\caption{Cost of Gas for Bidder's Operation (Simple Deposit Method)}
\label{gascost_bidder_naive}
\begin{tabular}{lll}
\hline
Operation & Used Gas & Fee  \\ \hline
Committing bid & 110,928  & 8.22 USD                         \\
Revealing bid & 83,119   & 6.16 USD                         \\ \hline    
\end{tabular}
\end{table}

\noindent 
As described in section \ref{estimate_gas}, the fee for the proposed protocol is about 19.65 USD (265,204 gas). 
Thus, the increased fee for hiding the maximum bidding price by the proposed method is about 14.37 USD (194,047 gas), and is about 37\% more expensive than that of the simple deposit method. 
Note that we use 1.56 USD as a normal transfer of Ethereum costs in this paper.

Next, we compare the fees for the operation of the open-bid auction.
As in the deposit method, we implemented the open-bid smart contract without changing the smart contract structure implemented in the proposed protocol as much as possible.
Table \ref{gascost_open} shows the gas used and fee for operations performed by the bidder.
The fee for bidding is approximately 5.27 USD, which is lower than the fee for the sealed-bid auction. 
However, multiple bids are generally placed in open-bid auctions. As a result, the sealed bid auction, which requires only one operation, may lower the fees.

\begin{table}[h]
\centering
\caption{Cost of Gas for Bidder's Operation (Open-Bid)}
\label{gascost_open}
\begin{tabular}{lll}
\hline
Operation           & Used gas & Fee  \\ \hline
Bidding & 71,137   & 5.27 USD                         \\ \hline    
\end{tabular}
\end{table}

\section{Analyses of Assumptions}

\subsection{How Much Bidding Transactions are Hidden}
\label{Tran_IND}
In this section, we consider how much bidding transactions are hidden. 
Because the one-time address is issued by a 256-bit salt which is known only by the bidder, it is difficult to guess the one-time address by others.
In addition, a bidding transaction to a one-time address and usual transactions in Ethereum are indistinguishable because a one-time contract has not been deployed yet. 
If malicious bidders want to find the bidding transaction, they can infer the one-time address from the following properties.

\smallskip
\begin{enumerate}
  \item Addresses that do not transfer ETH to other addresses during the bidding phase.
  \item Addresses that received ETH for the first time during the bidding phase.
\end{enumerate}

\smallskip
\noindent 
The first property is clear from the fact that the one-time contract has not been deployed yet in the bidding phase, and only the one-time contract, which is deployed in the revealing phase, can withdraw the funds of the one-time address.
The second property is also clear from the fact that the one-time address has not been used yet before the bidding phase. 
If there are only a few candidate addresses, malicious bidders may be able to guess the bidding price. 
To extract addresses that satisfy the two conditions, we used Google Cloud BigQuery, which has access to all data of Ethereum transactions.
The number of addresses with two conditions is shown in Table \ref{onetime_address}.

\begin{table}[h]
\centering
\caption{The Number of Addresses Satisfying the Two Conditions}
\label{onetime_address}
\begin{tabular}{lrrr}
\hline
                & Three Days & One Week & Two Weeks \\ \hline
0$\sim$0.1ETH   & 35,722      & 89,067          & 173,348          \\
0.1$\sim$0.5ETH & 4,458         & 10,229           & 18,668           \\
0.5$\sim$1ETH   & 1,286          & 2,944            & 5,368           \\
1$\sim$10ETH    & 2,361          & 5,558            & 10,505           \\
10$\sim$50ETH   & 573           & 1,193            & 2,196            \\
50$\sim$100ETH  & 94            & 198             & 347             \\
100$\sim$    & 168           & 329             & 558             \\\hline
\end{tabular}
\end{table}

\begin{figure*}[tb]
\centering
\includegraphics[width=0.9\linewidth]{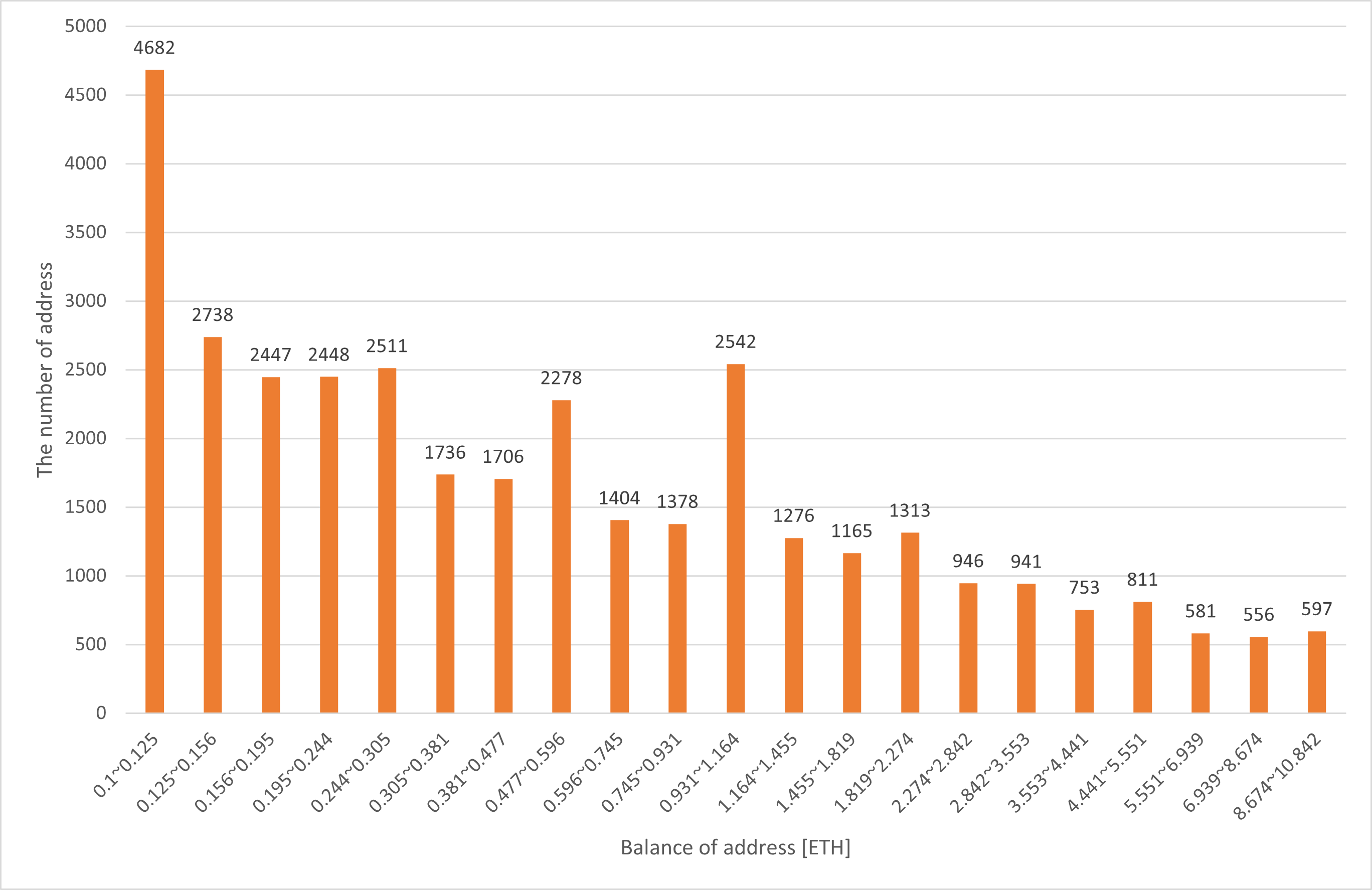}
\caption{The Number of Addresses Meeting the Two Conditions (0.1 ETH to 10 ETH)}
\label{fig:onetime}
\end{figure*}

We consider cases of three days (May 29 to 31, 2023), one week (May 25 to 31, 2023), and two weeks (May 18 to 31, 2023) as the bidding phase period. 
Assuming that the bidding price in the auction is 0.1 to 0.5 ETH, Table \ref{onetime_address} shows that there are 4,458 possible addresses in the three days case, 10,229 addresses in the one week case, and 18,668 addresses in the two weeks case.
Because the median price of the NFTs is approximately \$300 (0.16 ETH as of June 1, 2023), as described in Section \ref{background}, this assumption is reasonable. 
Figure \ref{fig:onetime} shows the number of addresses with balance between 0.1 ETH and 10 ETH during two weeks (May 18 to 31, 2023).
We remark that in many auctions, bidders know the approximate market price of an auctioned item, and can expect the range of bidding price, i.e., it does not become much higher/lower than the market price. For example, when an item is sold 0.125 ETH market price, then it seems reasonable that the winning bid is between 0.1 ETH and 0.15 ETH. Thus, we described the width of the graph in 1.25x intervals (between 0.1 and 0.125, between 0.125 and 0.156, and so on) that seems appropriately treat the above situation. Because all these addresses can be regarded as one-time addresses, it is clear that a longer bidding phase period is effective in hiding more one-time addresses.

\subsection{How Much the Maximum Bidding Price is Hidden}
\label{MAX_HIDE}

In the proposed protocol, the maximum bidding price is less than or equal to the maximum balance of all addresses.
However, many transactions for a relatively high ETH are transferred every day on Ethereum. For example, when we assume the bidding phase to be two weeks (May 18 to 31, 2023), the maximum balance of all addresses is 29,565 ETH.
It is potentially difficult to determine which transaction is for the maximum bidding price. 
On the other hand, in the case of a sealed-bid auction with the simple deposit method, the maximum bidding price is revealed. 

\section{Conclusion}

In this paper, we proposed a sealed-bid action protocol providing the fund binding property. We introduced transactions for one-time addresses and employed DECO protocol. We implemented our protocol and compared the protocol with the simple deposit method. Although we have discussed how much bidding transactions are hidden, there is a room for argument on this point. 
For example, bidders may be able to estimate a rough maximum bidding price from transactions that increased after the bidding phase. Alternatively, some statistical analyses may distinguish or identify transactions for one-time addresses or reveal the maximum bidding price. Thus, the statistical analysis of these transactions could be conducted in a future work. We expect that our blockchain-oriented technique could be a stepping stone for hiding transactions.

\medskip
\noindent\textbf{Acknowledgment}: This work was supported by JSPS KAKENHI Grant Numbers JP21K11897 and JP22H03588.

%


\end{document}